\documentclass[aps,prl,twocolumn,showpacs,superscriptaddress,longbibliography,floatfix,nofootinbib]{revtex4-2}

\usepackage{amsmath, amsfonts, amssymb, bm}
\usepackage{mathtools}
\usepackage[pdftex]{graphicx}
\usepackage{import}
\usepackage{ulem}
\usepackage{scalerel}
\usepackage{nccmath}
\usepackage{empheq}

\usepackage[breaklinks,hypertexnames=false]{hyperref}
\hypersetup{colorlinks,linkcolor={magenta},citecolor={blue},urlcolor={blue}} 

\usepackage{braket}
\usepackage{xcolor}

\usepackage[capitalize]{cleveref}

\usepackage{glossaries}
\glsdisablehyper

\newacronym{mbs}{MBS}{Majorana bound state}
\newacronym{mzm}{MZM}{Majorana zero mode}
\newacronym{abs}{ABS}{Andreev bound state}
\newacronym{fi}{FI}{ferromagnetic insulator}
\newacronym{jj}{JJ}{Josephson junction}
\newacronym{cpr}{CPR}{current-phase relation}

\usepackage[margin=0.9in]{geometry}
\usepackage[utf8]{inputenc}
\usepackage[inline,shortlabels]{enumitem}
\usepackage{longtable}
\usepackage[caption=false]{subfig}
\usepackage{multirow}

\usepackage{sidecap,tikz}
\definecolor{lime}{HTML}{A6CE39}
\DeclareRobustCommand{\orcidicon}{\hspace{-1mm}
	\begin{tikzpicture}
		\draw[lime, fill=lime] (0,0) 
		circle [radius=0.16] 
		node[white] {{\fontfamily{qag}\selectfont \tiny \,ID}};
		\draw[white, fill=white] (-0.0525,0.095) 
		circle [radius=0.007];
	\end{tikzpicture}
	\hspace{-3mm}
}
\foreach \x in {A, ..., Z}{\expandafter\xdef\csname orcid\x\endcsname{\noexpand\href{https://orcid.org/\csname orcidauthor\x\endcsname}
		{\noexpand\orcidicon}}
}

\def\be{\begin{equation}}
\def\ee{\end{equation}}
\def\bea{\begin{eqnarray}}
\def\eea{\end{eqnarray}}
\def\bmat{\begin{pmatrix}}
\def\emat{\end{pmatrix}}
\def\bs{\begin{split}}
\def\es{\end{split}}

\def\~{$\approx$}

\def\dag{\dagger}

\newcommand{\up}{\uparrow}
\newcommand{\dw}{\downarrow}

\newcommand{\PRLsep}{\noindent\makebox[\linewidth]{\resizebox{0.3333\linewidth}{1pt}{$\bullet$}}\bigskip}

\let\Re\undefined
\let\Im\undefined
\DeclareMathOperator{\Re}{\mathfrak{Re}}
\DeclareMathOperator{\Im}{\mathfrak{Im}}
\DeclareMathOperator{\diag}{diag}

\begin{document}

\title{Characterizing local Majorana properties using Andreev states}

\author{M. Alvarado\orcidA{}}
\affiliation{ 
Instituto de Ciencia de Materiales de Madrid (ICMM), Consejo Superior de Investigaciones Cient{\'i}ficas (CSIC), Sor Juana In{\'e}s de la Cruz 3, 28049 Madrid, Spain}

\author{A. Levy Yeyati\orcidD{}}
\affiliation{ 
Departamento de F{\'i}sica Te{\'o}rica de la Materia Condensada, Condensed Matter Physics Center (IFIMAC) and Instituto Nicol{\'a}s Cabrera, Universidad Aut{\'o}noma de Madrid, 28049 Madrid, Spain}

\author{Ram{\'o}n Aguado\orcidC{}}
\affiliation{ 
Instituto de Ciencia de Materiales de Madrid (ICMM), Consejo Superior de Investigaciones Cient{\'i}ficas (CSIC), Sor Juana In{\'e}s de la Cruz 3, 28049 Madrid, Spain}

\author{R. Seoane Souto\orcidB{}}
\affiliation{ 
Instituto de Ciencia de Materiales de Madrid (ICMM), Consejo Superior de Investigaciones Cient{\'i}ficas (CSIC), Sor Juana In{\'e}s de la Cruz 3, 28049 Madrid, Spain}

\date{\today}

\begin{abstract}
We propose using Andreev bound states (ABS) as spectroscopic probes to characterize Majorana zero modes (MZMs) in quantum-dot based minimal Kitaev chains. Specifically, we show that tunneling conductance measurements with a superconducting probe hosting an ABS reveal four subgap peaks whose voltage positions and relative heights enable extraction of the MZM energy splitting and Bogoliubov-de Gennes coherence factors. This provides direct access to zero‑splitting regimes and to the local Majorana polarization -- a measure of the Majorana character. The method is compatible with existing experimental architectures and remains robust in extended chains.
\end{abstract}

\maketitle

{\it Introduction ---} 
Majorana zero modes (MZMs) in topological superconductors~\cite{Wilczek2009, Alicea2012, Leijnse2012_b, Aguado2017, Beenakker2020, Tanaka2024}
are key building blocks of topological qubits that are intrinsically protected against local noise~\cite{Nayak2008, Sarma2015, Lahtinen2017, Aguado2020}. In recent years, quantum dot-based Kitaev chains (KCs)~\cite{Kitaev2001} have emerged as a promising and highly tunable platform for exploring this physics~\cite{Leijnse2012, Sau2012, Fulga2013, Leijnse2024}, predicted to host MZMs at discrete points in parameter space~\cite{Wang2022, Wang2023, Bordin2022, Liu2022, Dvir2023, Bordin2023, Haaf2023, Bordin2024, Bordin2025, Bordin2025b, vanloo2025}. So far, reported signatures of MZMs relied mostly on local and non-local transport using normal probes, that offer limited resolution of the underlying wavefunction structure, as they predominantly capture broadened tunneling features~\cite{Parity}.

On the other hand, other subgap states have gained renewed attention, not only for their fundamental role in hybrid superconducting devices~\cite{SeoaneAguado2024}, including novel qubits~\cite{Janvier2015,Hays2021, vanHeck2023, PitaVidal2023, Pita2025}, but also as precise and versatile probes. This second aspect has been mostly explored using scanning tunneling microscopy (STM) with superconducting (SC) tips, which enable detailed wavefunction analysis of, {\it e.g.}, Yu-Shiba-Rusinov (YSR) states~\cite{Yu1965, Shiba1968, Rusinov1969}. This technique exploits particle-hole asymmetry in the subgap conductance peaks to extract Bogoliubov-de Gennes (BdG) coherence factors~\cite{Ruby2015, Peng2015, Cuevas2020, Cuevas2020b, Perrin2022}. Similar concepts are now beginning to be explored in mesoscopic SC devices~\cite{Gorm2022, Bordin2025c}, although, for the best of our knowledge, they are underutilized for studying MZMs. 

In this Letter, we bridge this gap by demonstrating that Andreev bound states (ABSs) enable the extraction of the energy and the local BdG coherence factors of emergent low-energy states in KCs. In the weak-coupling regime, the differential conductance exhibits four subgap peaks, with heights $\alpha_\pm$ and  $\beta_\pm$. The voltage condition for these peaks directly relate to the energies of both subgap states, while the peak heights encode the corresponding BdG coherence factors. Specifically, two ratios formed from the four peak heights {\it are sufficient} to determine the relative BdG amplitudes of the probing ABS ($|u_A|$, $|v_A|$) and the KC ($|u_B|$, $|v_B|$) subgap states,
\be \label{coeff_charge}
\xi_1=\sqrt{\rule{0pt}{3.2ex} \frac{\alpha_- \beta_-}{\alpha_+ \beta_+}} = \frac{|u_A|^2}{|v_A|^2} \, , \quad \xi_2=\sqrt{\rule{0pt}{3.2ex} \frac{\alpha_+ \beta_-}{\alpha_- \beta_+}} = \frac{|u_B|^2}{|v_B|^2} \, .
\ee 
This method allows for extracting the local SC charge, defined as $q = |u|^2-|v|^2$, Fig.~\ref{fig1}(a), which vanishes when $|u|=|v|$, signaling an equal electron-hole superposition. Unlike alternative approaches based on non-local normal conductance\cite{Schindele2014, Gramich2017, Flensberg2018, Flensberg2020, Flensberg2020b, Lange2022}, our technique provides direct access to the local BdG coefficients ratio, directly tied to the degree of local Majorana protection. Therefore, spectroscopy using an ABS grants access to the energy splitting and enables the extraction of the Majorana polarization~\cite{Bena2015, Bena2016, Glodzik2020, Aksenov2020, Souto2022, Leijnse2024, Ola2024} (MP), thereby offering a clear diagnostic of sweet spot conditions~\cite{Souto2022, Seoane2023,Tsintzis2024}. Furthermore, the discrete ABS spectrum yields high resolution beyond that of conventional normal or even SC leads~\cite{Cuevas2020b}.

Since artificial KCs are based on quantum dots coupled via ABSs, we envision that some of these ABSs can be repurposed as built-in probes to characterize local Majorana properties of low-energy states and tune the device. This {\it in situ} approach is compatible with existing experimental setups and does not require additional ingredients~\cite{ Bordin2023, Bordin2024, Bordin2025, Bordin2025b, Bordin2025c}. The method remains effective even in extended chains and is inherently scalable, opening new pathways for optimizing experimental protocols and advancing the study of Majorana physics in longer architectures.

\begin{figure*}[t!]
\centering
\includegraphics[width=\textwidth]{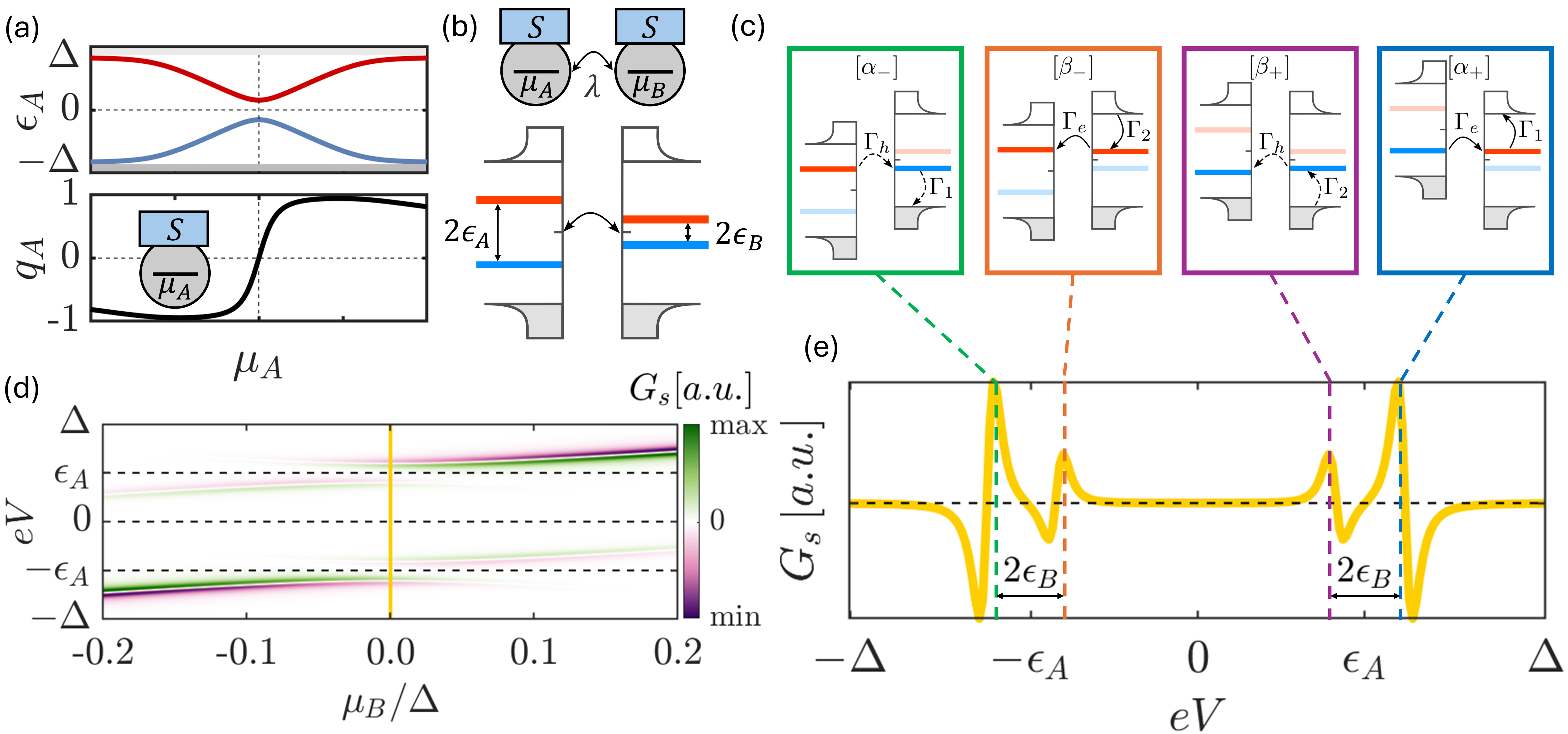}
\caption{(a) ABS behavior in a superconductor-quantum dot (S-QD) junction and the corresponding SC charge as a function of the QD chemical potential $\mu_A$.
(b) Schematic representation of a generic ABS-ABS tunneling scheme.
(c) Diagrams of the on-resonance dominant transport mechanisms. 
(d) SC differential conductance between generic ABSs as a function of the target subgap state detuning $\mu_B$, showing both the dispersion of conductance peaks and the emergence of a peak height asymmetry. We set $\mu_A=0$ to avoid extra asymmetries coming from the probe.
(e) Symmetric conductance cut along the yellow line at $\mu_B = 0$ showing the distinctive four peaks associated to different tunneling processes, and with negative differential conductance values characteristic of the weak-coupling regime.}
\label{fig1}
\end{figure*}

{\it Physical Concept ---} 
We first study the ABS spectroscopy of a generic SC subgap state. The latter is described by the Bogoliubov quasiparticle operator $\gamma_B = u_B \, c + v_B \, c^\dag$ with energy $\epsilon_B$, while the probing ABS has energy $\epsilon_A$. 
Both subgap states are coupled through $\lambda$, see Fig.~\ref{fig1}(b). 
In this section, we treat both target and probing ABSs using a minimal description based on the large SC gap limit~\cite{Bauer2007} ({\it i.e.}, $\Delta \rightarrow \infty$), with coherence factors
\begin{gather}\label{Eq_atomiclimit}
|u_\nu|^2=\frac{\epsilon_\nu+\mu_\nu}{2\epsilon_\nu} \, , \qquad |v_\nu|^2=\frac{\epsilon_\nu-\mu_\nu}{2\epsilon_\nu} \, , \nonumber \\
\epsilon_\nu = \sqrt{\Delta_\nu^2+\mu_\nu^2} \, ,
\end{gather}
where $\Delta_\nu$ is the effective gap and $\mu_\nu$ is the chemical potential ($\nu=A,B$).

In the weak-coupling regime, transport between subgap states is dominated by single-particle tunneling~\cite{Ruby2015, Cuevas2020}. This results in four conductance peaks, corresponding to individual tunneling processes, Figs.~\ref{fig1}(c,e). These processes are characterized by frequency-dependent tunneling~\cite{Ruby2015, Peng2015, Gorm2022}
\bea
\Gamma_{e} &=& 2\pi\lambda^2 \, |u_B|^2 \, \rho_{A,e}(\omega-eV) \, , \nonumber \\
\Gamma_{h} &=& 2\pi\lambda^2 \, |v_B|^2 \, \rho_{A,h}(\omega+eV) \, , 
\eea
where the $|u_B|^2$ and $|v_B|^2$ prefactors reflect the probability amplitudes for electron- or hole-like tunneling into the target subgap state, and $\rho_{A,e/h}(\omega)$ are the Nambu entries of the spectral density at the probe and depend on the BdG coherence factors of the ABS, $|u_A|$ and $|v_A|$, following

\be 
\rho_{A,e/h} = \frac{\Lambda}{\pi} \Bigg [ \frac{|u_A|^2}{(\omega \mp \epsilon_A)^2+\Lambda^2} + \frac{|v_A|^2}{(\omega \pm \epsilon_A)^2+\Lambda^2} \Bigg ]\, ,
\ee
being $\Lambda$ a broadening term of the probing ABS~\cite{Dynes1978}, cf. the Supplemental Material~\cite{SM} for a comprehensive derivation. As such, both $\Gamma_{e}$ and $\Gamma_{h}$ carry information about the internal structure of the subgap states involved in the tunneling. 
 
These single-particle processes change the occupation of the target subgap state and require relaxation, characterized by the rates $\Gamma_1$ (emptying) and $\Gamma_2$ (filling)~\cite{ALY2014, Ruby2015}. To sustain a steady single-particle current, relaxation must occur much faster than tunneling~\cite{Peng2015, Ruby2015, Alvarado2024} ($\Gamma_{1,2} \gg \Gamma_{e/h}$). In realistic SC devices, processes such as quasiparticle poisoning, phonon emission, and photon absorption can affect subgap occupations~\cite{ALY2014, Ruby2015}, but we assume here the simplest situation of thermal broadening given by $\Gamma_1 = \Gamma_t\, [1 - n_F(\omega)] $  and $\Gamma_2 = \Gamma_t\, n_F(\omega)$ respectively, with $n_F(\omega)$ being Fermi functions, cf. the Supplemental Material~\cite{SM} for a discussion on the parameter regimes. In contrast, resonant Andreev reflection, is a higher-order process that does not change occupancy and becomes dominant at stronger couplings~\cite{Cuevas2020} ($\Gamma_{1,2} \ll \Gamma_{e/h}$).

Figures~\ref{fig1}(d,e) shows the differential conductance $G_s = \partial I_s / \partial V$, where each peak corresponds to a different single-particle tunneling process, occurring at the voltage thresholds $[\alpha_\pm] = \pm(\epsilon_A + \epsilon_B)$ and $[\beta_\pm] = \pm(\epsilon_A - \epsilon_B)$. These peaks are symmetrically positioned in bias around $e|V|=\epsilon_A$ and are separated by an energy difference of $2\epsilon_B$. By tuning the chemical potential $\mu_B$, we control the energy of the target subgap state and its BdG coefficients. This manifests as a shift in the peak positions and the peak heights, since the latter depend sensitively on the BdG wavefunction amplitudes through $\Gamma_{e/h}$. In this way, the subgap conductance spectrum provides a direct spectroscopic fingerprint of the coherence factors that characterize the target subgap state, but also the probing ABS in the junction.

The dominant conductance peaks $G_s([\alpha_\pm]) \to \alpha_\pm$ correspond to direct quasiparticle tunneling, while the secondary ones $G_s([\beta_\pm]) \to \beta_\pm$ result from thermally activated processes, see Fig.~\ref{fig1}(e) and, cf. the SM~\cite{SM}. For realistic conditions where $\epsilon_B > k_BT \gg \Gamma_t$~\cite{Flensberg2020, Perrin2022}, all peaks remain well resolved and dominated by thermal broadening, which ultimately sets the resolution limit of the method. More specifically, the conductance peaks for positive bias thresholds satisfy
\bea \label{Eq_conductance}
G_s([\alpha_+]) \propto \frac{e^2 \lambda^2}{h} \frac{\Lambda \, \Gamma_{1} \, |v_A|^2 \, |u_B|^2 }{(\Gamma_t^2/4-\Lambda^2)^2} \, , \nonumber \\
G_s([\beta_+]) \propto \frac{e^2 \lambda^2}{h} \frac{\Lambda \, \Gamma_{2} \, |v_A|^2 \, |v_B|^2}{(\Gamma_t^2/4-\Lambda^2)^2} \, ;
\eea
with similar expressions at negative bias following 
the transformations $(v_A \rightarrow u_A)$, and $(u_B \leftrightarrow v_B)$. 

Importantly, by forming suitable ratios of the above conductance peaks, one can isolate and extract key physical parameters: the ratio of coherence factors for the probing ABS ($|u_A|/|v_A|$), for the target subgap state ($|u_B|/|v_B|$), and the ratio of relaxation rates ($\Gamma_1/\Gamma_2$), in a way that is independent of device-specific parameters such as $\Lambda$, and $\lambda$.

{\it Short Kitaev Chains ---} 
To illustrate the potential of the method, we now demonstrate that it can be used to extract relevant physical properties of the emergent subgap states in minimal KCs. We consider a chain coupled to a normal lead and an ABS, localized at a quantum dot (QD) strongly coupled to a mesoscopic SC lead~\cite{Bauer2007} ({\it i.e.}, it includes the quasiparticle continuum~\cite{Alvarado2024}). A dc bias is applied between the SC probe and the grounded system, ensuring that the measured current flows through the ABS into the chain, see Fig.~\ref{fig2}(a).

The KC is described by the spinful model in Ref.~\cite{Dominguez2016, Bordin2022, Souto2022, Liu2023, Miles2023, Souto2025},  where $\mu$ ($\mu_c$) is the on-site chemical potential for the QDs (SC region), which can be tuned by independent external gates, cf. the Supplemental Material~\cite{SM} for details on the model. Both, $\mu$ and $\mu_c$ modify the energy, the localization, and the BdG coefficients of the lowest-energy state, Fig.~\ref{fig2}(a), making KCs an ideal platform to probe our predictions. Here, we focus on the local MP that determines the Majorana character of the wavefunction~\cite{Bena2015, Bena2016, Glodzik2020, Aksenov2020}

\be \label{localMP}
{\cal P}_j = \frac{\sum_\sigma 2 \, u_{j \sigma} v_{j \sigma}}{\sum_\sigma \big|u_{j \sigma} \big|^2 + \big|v_{j \sigma} \big|^2} \, ,
\ee 
where $u_{j\sigma}$ and $v_{j\sigma}$ are the spin-resolved BdG amplitudes of the lowest-energy state at site $j$.  
The minimal KC couples with a tunneling amplitude $\lambda$ to a proximitized QD with chemical potential $\mu_A$, that hosts an ABS. The conductance is obtained numerically, using the recursive non-equilibrium Green's functions developed in Refs.~\cite{Cuevas1996, Zazunov2016, Alvarado2020, Alvarado2022}, cf. the Supplemental Material~\cite{SM} for details of the calculation. 

\begin{figure}[t!]
\centering
\includegraphics[width=\columnwidth]{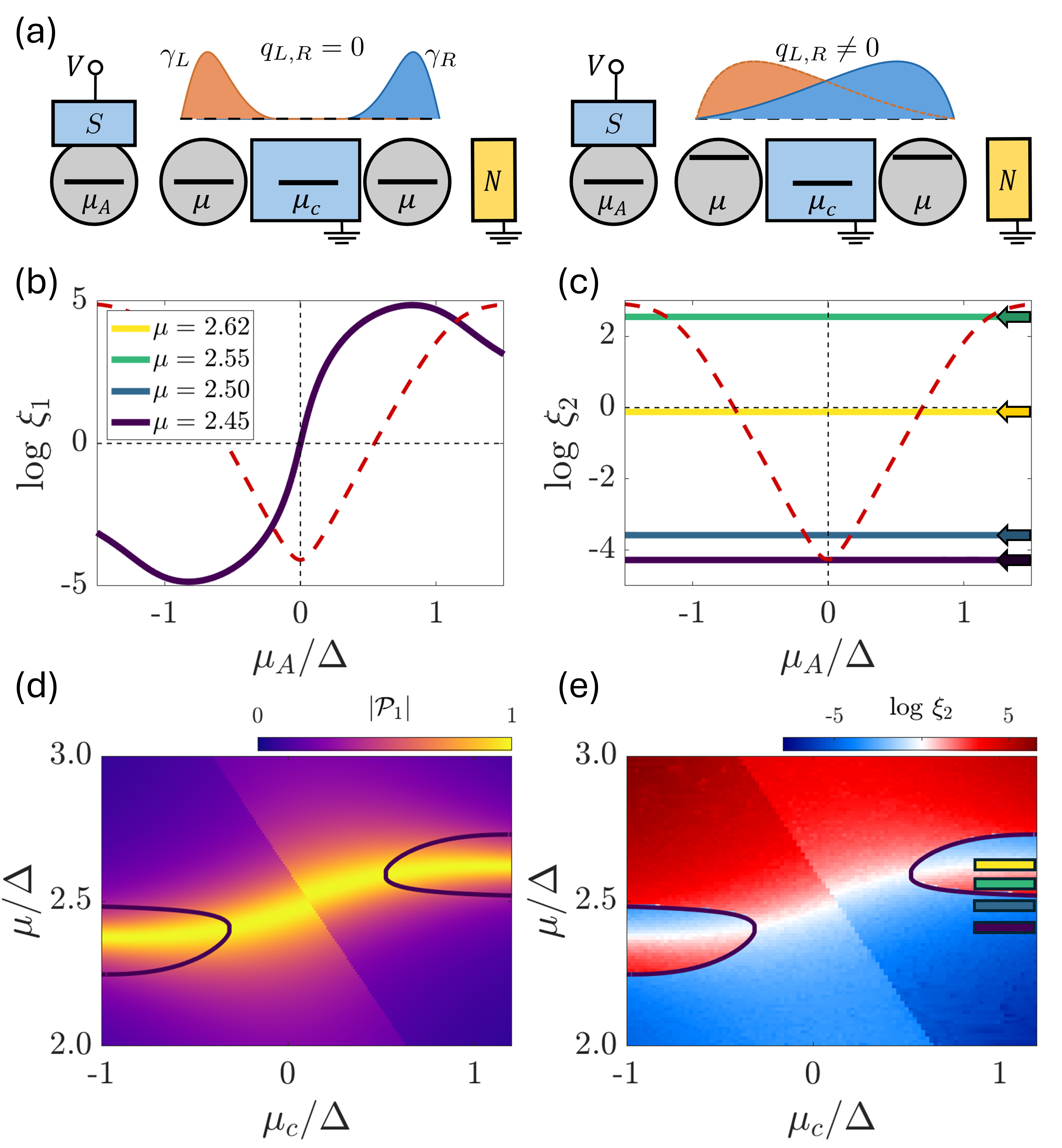}
\caption{(a) Illustration of the MZM localization in a minimal KC at (left) and away (right) from the Majorana sweet spot. (b, c) Evolution of $\log\,\xi_1$ and $\log\,\xi_2$ in Eq.\eqref{coeff_charge} as a function of the ABS detuning $\mu_A$, for several values of the outer QDs chemical potential $\mu = 2.45$, $2.5$, $2.55$, $2.62$, with $\mu_c = 1.2$, all in $\Delta$ units. Red dashed line marks the dispersion of the probing ABS. Arrows in panel (c) indicate the $|u_B|^2/|v_B|^2$ values obtained from Hamiltonian diagonalization. (d, e)  Phase diagram of the minimal chain as a function of $\mu$ and $\mu_c$. Panel (d) shows the calculated local MP at QD1 ($|{\cal P}_1|$), while (e) shows the corresponding $\log , \xi_2$ extracted from transport. Dark blue line denotes the analytical contour for $\epsilon_B = 0$, cf. the SM~\cite{SM}.
}
\label{fig2}
\end{figure}

Figures~\ref{fig2}(b,c) shows the evolution of the different BdG coherence factors extracted from transport as a function of $\mu_A$, for various values of $\mu$ in a minimal KC.
It is worth emphasizing that $\log \, \xi_1$  reproduces qualitatively the charge of the probing ABS and, in the limit of weak coupling, remains independent of the KC parameters, see Fig.~\ref{fig2}(b) where $\log\,\xi_1$ is represented for different values of $\mu$. Deviations from this behavior can thus serve as an experimental diagnostic for the sensing setup.

We next analyze the $\log\,\xi_2$ ratio, Fig.~\ref{fig2}(c). This ratio shows robust behavior largely unaffected by the dispersion of the ABS and excellent agreement with the BdG coefficients obtained by exact diagonalization, see arrows. The agreement is specially good in regions where $|u_B| \sim |v_B|$ and coupling strengths ($\lambda \ll\sqrt{\Gamma_t}$ in $\Delta$ units). 
In particular, the ABS's sensitivity is maximized
when operating at regions where the ABS charge is maximal, $|q_A| \approx 1$ ({\it i.e.}, near the peaks of $|\log \, \xi_1|$) ensures an optimal regime for probing the subgap states in the KC. It is important to note that the analytical approach breaks down in regions where $|u_A| \sim |v_A|$, as additional contributions to the current become significant, cf. the SM~\cite{SM}.

Figures~\ref{fig2}(d,e) show the local MP and $\log \, \xi_2$ when detuning both $\mu$ and $\mu_c$ in the chain. Points with high local MP, {\it i.e.}, $|{\cal P}_1| = 1$, correlate with $\log \, \xi_2 = 0$, which, in turn, indicates zero local charge $|u_B| = |v_B|$. We note that $\log \xi_2$ features two kind of sign changes: continuously, crossing zero, or abruptly, jumping from positive to negative values.
The continuous behavior reflects a smooth crossover between $u$- and $v$-dominated character, associated with high-MP behavior. In contrast, sharp jumps are due to a crossing of the KCs low-energy states, where the superconducting charge peaks as $\epsilon_B$ crosses zero~\cite{Schindele2014, Gramich2017, Flensberg2018, Flensberg2020, Flensberg2020b}, marked with solid line. This difference allows to search for zero-energy states and high MP.

{\it Longer Chains ---} 
To further assess the scalability and robustness of our approach, we extend the analysis to longer KCs~\cite{Miles2023, Liu2025, Luethi2025,Dourado2025} with 5- and 10-sites. Figure~\ref{fig3} demonstrates that the spectroscopic coefficient $\log \, \xi_2$ closely tracks regions of maximum local MP, even when it is nearly saturated, capturing subtle variations across the phase diagram. The coherence factors extracted from transport remain in strong agreement with those obtained via Hamiltonian diagonalization, despite the increased number of QDs~\cite{SM}.

Nonetheless, several practical considerations arise beyond the ideal limit. First, the method's resolution is ultimately limited by the intrinsic broadening of the conductance peaks, governed by the relaxation rate $\Gamma_t$. Therefore, the minimal resolvable energy is set by $\Gamma_t$, which becomes critical when the gap between states is small. Second, voltage fluctuations can shift peak positions, inducing errors on the peak height measurements -- yet accuracy is maintained if the peaks deviations stay below the broadening scale.

\begin{figure}[t!]
\centering
\includegraphics[width=\columnwidth]{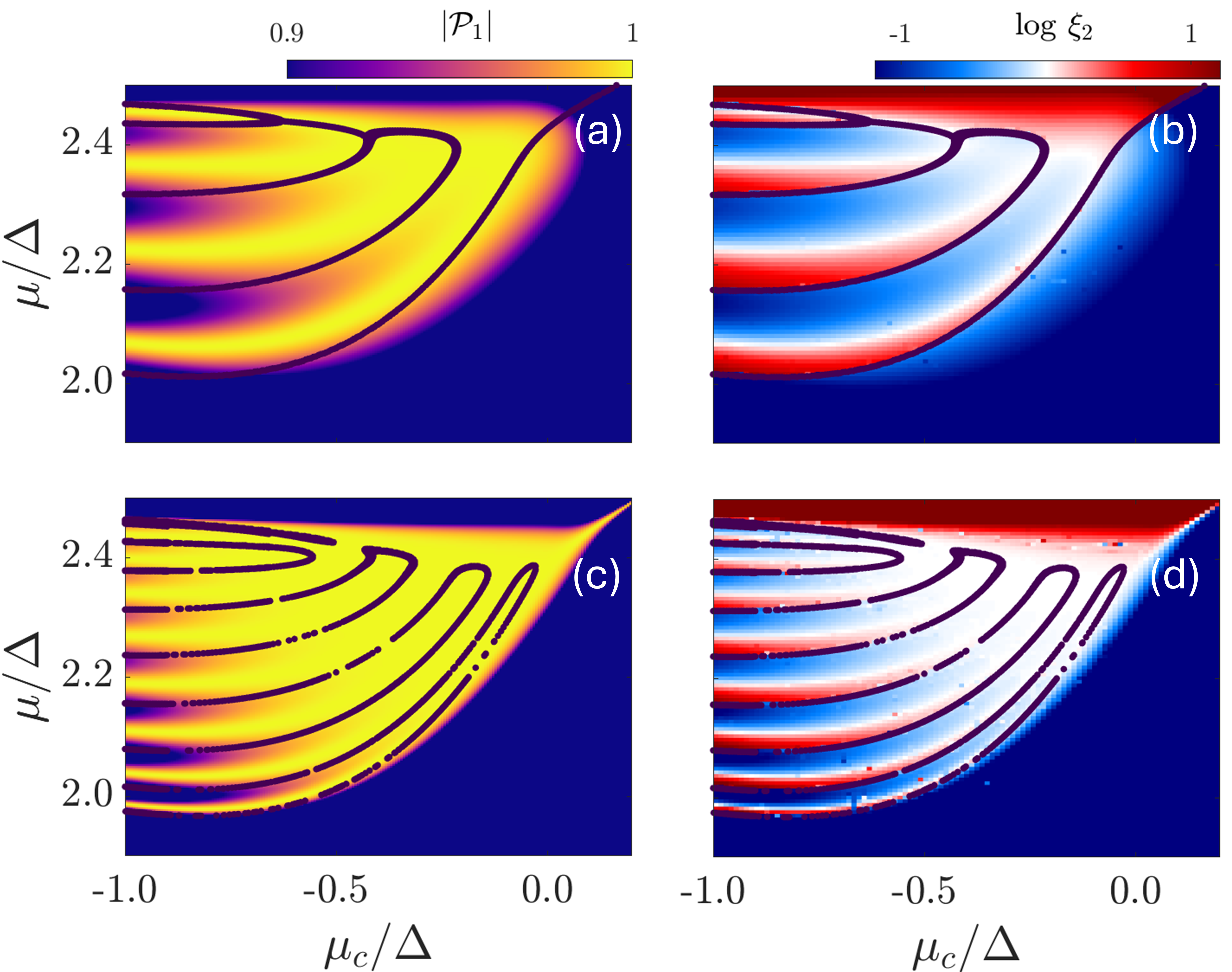}
\caption{
Phase diagram for chains with $n=5$ (top panels) and $n=10$ (bottom panels) sites, as a function of the chemical potentials $\mu$ and $\mu_c$, all set to the same value.
Left column, Local MP at the boundary site, $|{\cal P}_1|$.
Right column, $\log \, \xi_2$ extracted from transport conductance. Dark blue line denotes the contour for $\epsilon_B = 0$.
}
\label{fig3}
\end{figure}

A further challenge is distinguishing true MZMs from quasi-Majoranas -- overlapping Majorana modes that emerge in smooth potential profiles~\cite{Prada2012, Kells2012, Liu2012, Fer2018,Vuik2019, Avila2019}. In the weak-coupling regime, a probing ABS treats quasi-Majoranas as two independent channels for transport. Therefore, the method based on the MP extraction through Eq.\eqref{coeff_charge} cannot differentiate them from localized MZMs. However, we expect that at a stronger coupling, the ABS can hybridize with the quasi-Majoranas~\cite{SM}, allowing to extract information about their spatial overlap. This approach generalizes earlier ideas from Refs.~\cite{Deng2016, Prada2017, Clarke2017, Seoane2023, Bordin2025d}, which suggest that QDs can be used as a spectroscopic tool.

On the other hand, local and non-local MP can differ in long chains~\cite{Ola2024, Tsintzis2024}, cf. the Supplemental Material~\cite{SM}. Our proposed spectroscopic method provides access to the local MP, relevant quantity when targeting local experiments, {\it e.g.}, a qubit based on MZMs in KCs~\cite{Tsintzis2024,Pan2025}. Thus, local MP provides a reliable signature for the presence of MZMs in Majorana experiments and for scalable architectures.

{\it Conclusions and Outlook ---} 
In this Letter, we have demonstrated that an Andreev bound state can be utilized as a sensitive probe for low-energy states appearing in quantum dot-based Kitaev chains. As an intrinsic component of these systems, it enables better energy resolution than other metallic or superconducting probes. The local conductance through the ABS allows to extract the Bogoliubov-de Gennes coherence factors of low-energy states. This approach identifies sweet spots with well-localized Majorana states by pinpointing regions of vanishing local charge and zero-energy splitting. Furthermore, we have successfully extended this methodology to longer chains.

Here, we focused on the weak coupling between the low-energy states appearing in KCs and Andreev bound states. The strong tunnel limit can provide further information about the robustness of the emergent states in KCs, eventually distinguishing Majorana from other trivial zero-energy states.

{\it Note Added ---} A related work by R. Dourado et al., Ref.~\cite{Dourado2025b}, submitted in parallel, proposes extracting the Majorana polarization using normal transport via local current measurements by changing the tunnel coupling to a lead.

{\it Acknowledgments ---} We thank A. Bordin and G. O. Steffensen for valuable comments and discussions. Work supported by the Horizon Europe Framework Program of the European Commission through the European Innovation Council Pathfinder Grant No. 101115315 (QuKiT), the Spanish Comunidad de Madrid (CM) ``Talento Program'' (Project No. 2022-T1/IND-24070), the Spanish Ministry of Science, Innovation, and Universities through Grants CEX2024-001445-S (Severo Ochoa Centres of Excellence program), PID2022-140552NA-I00, PID2021-125343NB-I00, and TED2021-130292B-C43 funded by MCIN/AEI/10.13039/501100011033, ``ERDF A way of making Europe'' and European Union Next Generation EU/PRTR. Support from the CSIC Interdisciplinary Thematic Platform (PTI+) on Quantum Technologies (PTI-QTEP+) is also acknowledged.




\providecommand{\noopsort}[1]{}\providecommand{\singleletter}[1]{#1}%
%


\onecolumngrid
\vspace{1mm}
\PRLsep

\newpage 

\vspace{5mm}
\centerline{\large\bfseries Supplemental Material to ``Characterizing local Majorana properties using} 
\centerline{\large\bfseries  Andreev states'' }
\vspace{5mm}

\twocolumngrid

\setcounter{equation}{0}
\renewcommand{\theequation}{S\arabic{equation}}

\setcounter{figure}{0}
\renewcommand{\thefigure}{S\arabic{figure}}

\title{Supplemental Material to ``Characterizing local Majorana properties using Andreev states''} 

\author{M. Alvarado\orcidA{}}
\affiliation{Instituto de Ciencia de Materiales de Madrid (ICMM), Consejo Superior de Investigaciones Cient{\'i}ficas (CSIC), Sor Juana In{\'e}s de la Cruz 3, 28049 Madrid, Spain}

\author{A. Levy Yeyati\orcidD{}}
\affiliation{Departamento de F{\'i}sica Te{\'o}rica de la Materia Condensada, Condensed Matter Physics Center (IFIMAC) and Instituto Nicol{\'a}s Cabrera, Universidad Aut{\'o}noma de Madrid, 28049 Madrid, Spain}

\author{Ram\'{o}n Aguado\orcidC{}}
\affiliation{Instituto de Ciencia de Materiales de Madrid (ICMM), Consejo Superior de Investigaciones Cient{\'i}ficas (CSIC), Sor Juana In{\'e}s de la Cruz 3, 28049 Madrid, Spain}

\author{R. Seoane Souto\orcidB{}}
\affiliation{Instituto de Ciencia de Materiales de Madrid (ICMM), Consejo Superior de Investigaciones Cient{\'i}ficas (CSIC), Sor Juana In{\'e}s de la Cruz 3, 28049 Madrid, Spain}

\date{\today}

\maketitle

\setcounter{equation}{0}
\renewcommand{\theequation}{S\arabic{equation}}

\setcounter{figure}{0}
\renewcommand{\thefigure}{S\arabic{figure}}

\section{Modelization and Parameters} \label{App_A}
In the main text we examine a Kitaev chain (KC) coupled to a normal lead (N) and an Andreev bound state (ABS), localized at a quantum dot (QD) strongly coupled to a superconducting (SC) lead. The Kitaev chain Hamiltonian $H_{kc}=H_{0}+H_T$ is described by the components~\cite{Dominguez2016, Bordin2022, Souto2022, Liu2023, Miles2023, Souto2025}
\bea\label{model_Hamiltonian}
H_{0} &=& \sum^L_{i} \sum_{\sigma} (\mu_{i}+s_\sigma \, V_{z,i}) \, n_{i,\sigma} + \nonumber \\ && U_i \, n_{i,\up} \, n_{i,\dw} +  \Delta_i \, c_{i,\up}^\dag c_{i,\dw}^\dag + H.c. \, , 
\eea
where $L$ is the total length of the KC and $\mu_{i}$ is the on-site chemical potential, which can be tuned by independent external gates, satisfying that for odd (even) sites $\mu_i = \mu$ ($\mu_i = \mu_c$). The Zeeman field along the $z$-axis is $V_{z,i}=V_z$, where $s_{\up, \dw}=\pm1$, and $n_{i,\sigma}=c_{i,\sigma}^\dag c_{i,\sigma}$ is the number operator, with $c_{i,\sigma}^\dag$ ($c_{i,\sigma}$) representing the particle creation (annihilation) local operators with spin $\sigma$. Only even-site QDs are strongly coupled to SCs, leading to the effective pairing $\Delta_i = \Delta$, where we have omitted the phase difference and assumed $V_{z,i}=0$ due to the strong renormalization of the g-factor of even QDs by the SC hybridization~\cite{Miles2023}. The tunneling between QDs is described by
\be 
H_T = \sum_{i}^{L-1}\sum_\sigma t \, c_{i+1,\sigma}^\dag c_{i,\sigma} + t_{so} \, s_\sigma \, c_{i+1,\sigma}^\dag c_{i,\bar{\sigma}} + H.c. \, ,
\ee 
where $t$ ($t_{so}$) is the spin-conserving (spin-flipping) hopping, with the spin-orbit field along the y-axis, and $\bar{\sigma}$ denotes the opposite spin to $\sigma$. The Coulomb repulsion $U_i$ is neglected in this work without loss of generality as it only induces a renormalization of the sweet spots. This model maps onto the KC when $(\Delta, V_z \gg t)$ and can be particularized for any number of sites. 

From the KC Hamiltonian, we extract the local Majorana polarization (MP) at each site following~\cite{Bena2015, Bena2016, Glodzik2020, Ola2024}

\be 
{\cal P}_j = \frac{\sum_\sigma 2 \, u_{j \sigma} \, v_{j \sigma}}{\sum_\sigma \big|u_{j \sigma} \big|^2 +  \big|v_{j \sigma} \big|^2} \, ,
\ee 
where $u_{j \sigma} \, , \, v_{j \sigma}$ are the electron and hole components with spin $\sigma$ of the eigenstate $\Psi_j$ of the lower energy state at $\epsilon_B$. From the local contributions it is possible to define a non-local MP~\cite{Bena2016, Ola2024} 
\be \label{Global_MP}
{\cal P} = \Bigg [\sum_{j=1}^{L/2} {\cal P}_j \Bigg] \, \Bigg[ \sum_{j=L/2+1}^{L} {\cal P}_j \Bigg]^* \, ,
\ee 
where ${\cal P} = - 1$ indicates that Majorana zero modes are true zero modes (MZMs) and do not overlap. 

The chain is coupled to a normal lead through the tunneling rate $\Gamma_n = \lambda_n^2/t_n$ and to an ABS via the coupling $\lambda$ to the proximitized QD with chemical potential $\mu_A$. The latter is coupled itself to a SC lead through the tunneling rate $\Gamma_s = \lambda_s^2/t_s$, with pairing $\Delta_s = \Delta$. Thus, $\lambda_s$ ($\lambda_n$) represents the coupling and $t_s$ ($t_n$) denotes the bandwidth for the SC (normal) lead. Unless otherwise stated, we use $\Delta=2\,t=10\,t_{so}=1$, $V_z=2.5 \, \Delta$, and $t_s=t_n=10 \, \Delta$. 

For conductance calculations, we use $k_BT = 0.1 \, \Delta$~\cite{Ruby2015, Perrin2022} and set $\mu_A=-0.8 \, \Delta$ ensuring that $|q_A|\approx1$, thereby avoiding the regime $|u_A|\sim|v_A|$ where our transport analysis would not be fully applicable.
The intrinsic broadening of the ABS and the MZM is set to $\Lambda = \Gamma_t \sim 10^{-3} - 10^{-4}  \cdot \Delta$~\cite{Ruby2015, Cuevas2020, Gorm2022}. Under optimal conditions, where $\epsilon_B > k_BT \gg \Gamma_t$~\cite{Perrin2022}, the thermal broadening sets the lower bound for resolvable energy splittings. The coupling to the different elements of the junction are $\lambda_s = \Delta$, and $\lambda_n \ll \sqrt{\Gamma_t}$ in $\Delta$ units to avoid extra contributions to the relaxation. 

Finally, it is required a sufficiently small coupling strength between subgap states to ensure that single-particle tunneling dominates, typically showing normal state conductance values $G_n<10^{-3} \cdot G_0$~\cite{Ruby2015, Cuevas2020} in units of the quantum of conductance $G_0=2e^2/h$, such $\lambda \propto\sqrt{G_n}$, and satisfying $\lambda \ll \sqrt{\Gamma_t}$ in $\Delta$ units. Therefore, we estimate $\lambda \lesssim 10^{-2}-10^{-4} \cdot \Delta$ as an upper end for the coupling. 
It should be noted that $\lambda_n$ could be used in experiments to increase the relaxation rate $\Gamma_t$ within the chain, thereby avoiding measurement regimes that require ultra-weak couplings -- albeit at the cost of reduced resolution.

\begin{figure}[t!]
\centering
\includegraphics[width=\columnwidth]{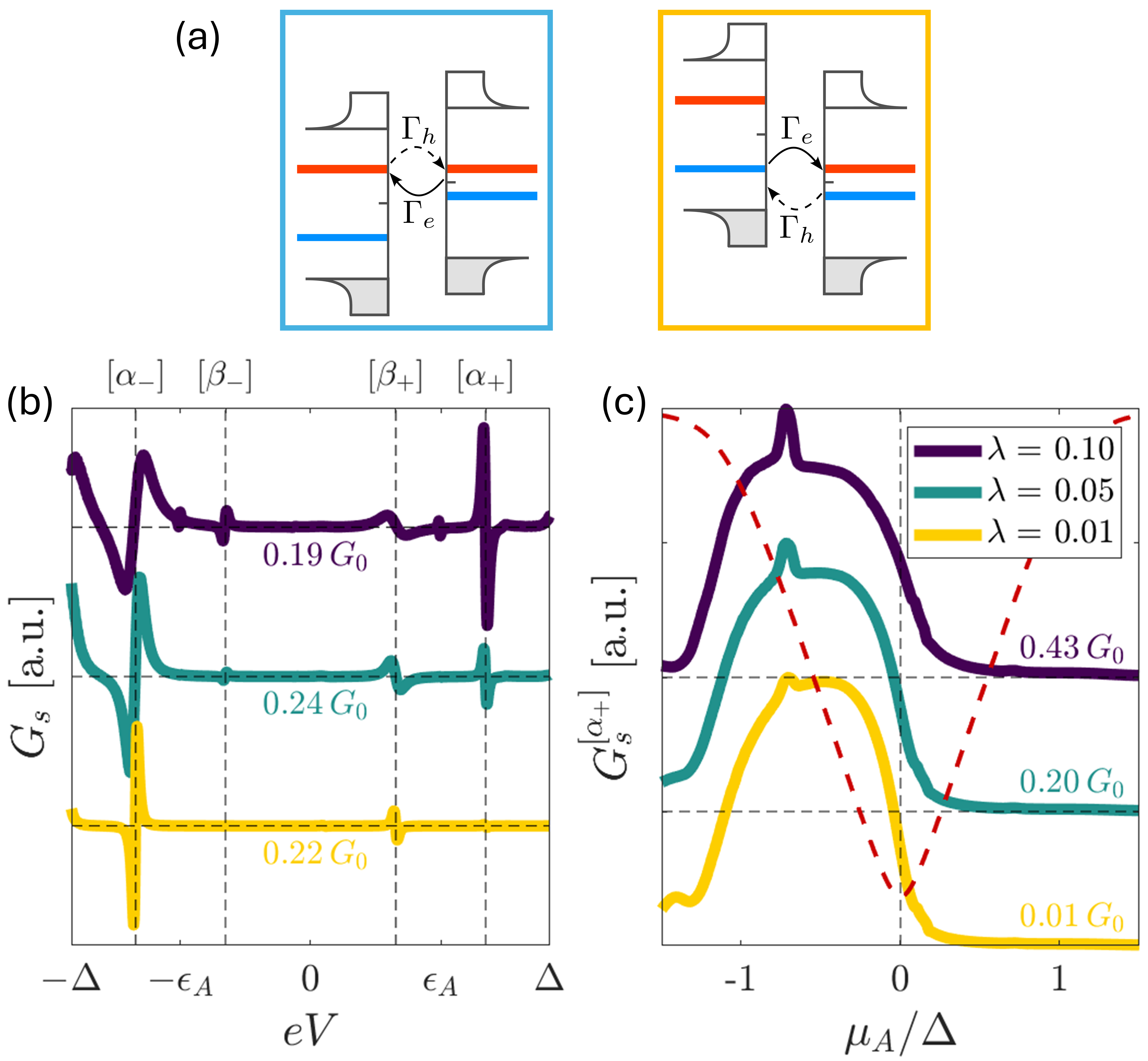}
\caption{(a) Schematics of resonant Andreev reflection processes occurring at voltages $[\alpha_-]$ and $[\alpha_+]$.
(b) SC differential conductance for various coupling strengths between the KC and the ABS, $\lambda/\Delta = 0.01$, $0.05$, and $0.1$, normalized to the main peak value, with $G_0 = 2e^2/h$. 
We set , and $\mu_A=0$ to avoid extra asymmetries coming from the probe. (c) Conductance at $[\alpha_+]$ as a function of the ABS detuning $\mu_A$, and normalized to the resonance value at $\mu_A/\Delta \approx -0.7$. Red dashed line indicates the dispersion of the probing ABS. We use $\mu/\Delta = 2$, $\mu_c/\Delta = -1$,  and the broadening $\Gamma_t = \Lambda = 5 \times 10^{-3} \cdot \Delta$, chosen for visualization clarity.}
\label{fig4}
\end{figure}

\section{Conductance phenomenology} \label{App_B}

In the main text, we focused on the linear conductance regime, corresponding to the weak coupling between the Andreev bound state and the Kitaev chain. In this section, we go beyond this limit and discuss of some features that arise in the differential conductance due to stronger couplings and higher-order tunneling processes. These effects  illustrate the richness of the underlying transport phenomenology and help delineate the regime of validity for our method.

To describe transport across the device we use the non-equilibrium Green's function formalism~\cite{Cuevas1996, Zazunov2016, Alvarado2020, Alvarado2022}, that allows us to consider arbitrary coupling strengths between the KC and the leads, cf. the sections below for further technical details. However, our formalism does not describe the multiple Andreev reflection (MAR) processes, that lead to subgap conductance peaks at voltage bias $V=2\Delta$/n, being $n$ an integer number. As a complement to the main text, Fig.~\ref{fig4}(a) schematically shows the  resonant Andreev reflection processes between the ABS and the KC~\cite{Cuevas2020}, whose amplitude is proportional to $\Gamma_{e} \, \Gamma_{h} \propto \lambda^4$. These contributions become prominent at strong tunnel couplings ($\lambda \gtrsim \sqrt{\Gamma_t}$ in $\Delta$ units) and in the zero-temperature limit ($T \to 0$), since Andreev processes do not alter the KC's occupation and thus do not require thermal relaxation.

Figure~\ref{fig4}(b) shows the conductance as a function of the applied bias voltage when going from weak to strong coupling. The figure shows that the conductance peaks broaden when going toward the strong coupling regime. In particular, the dips disappear in the strong coupling regime (not shown)~\cite{Peng2015, Ruby2015, Cuevas2020b, Gorm2022, Alvarado2024}. Moreover, we certify an inversion of the relative heights of the $\alpha_\pm$ peaks as the coupling strength increases: for weak coupling $\alpha_+<\alpha_-$, while for stronger coupling, the relation reverses to $\alpha_+>\alpha_-$. This inversion signals the crossover between linear and sublinear regime ({\it i.e.}, when $\Gamma_t \lesssim \Gamma_{e,h}$), and is known to be sensitive to temperature~\cite{Ruby2015, Gorm2022}.

\begin{figure}[b!]
\centering
\includegraphics[width=\columnwidth]{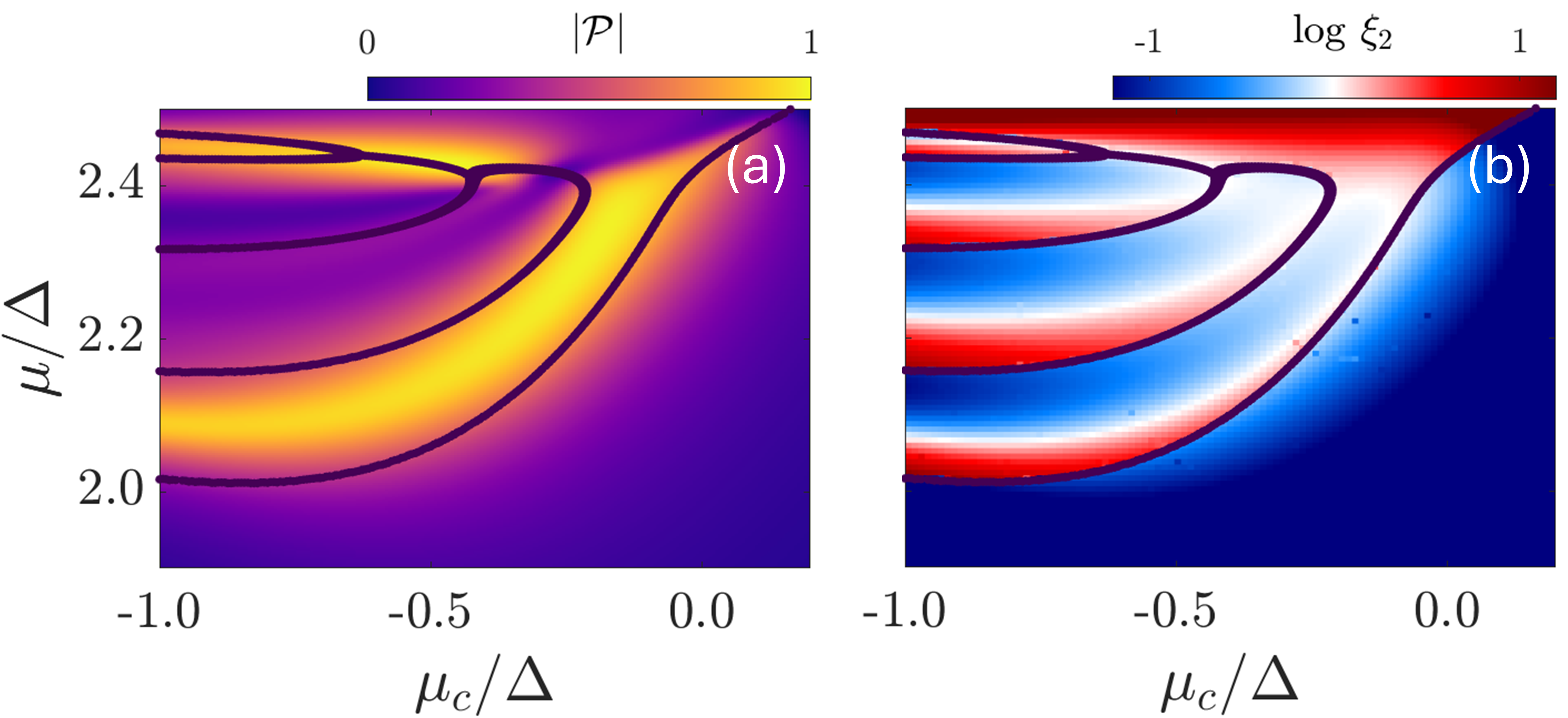}
\caption{Phase diagram of a 5‑site Kitaev. (a) Global Majorana polarization $|{\cal P}|$, reflecting the non-local topological character of the ground state.
(b) Map of the coefficient $\log , \xi_2$ extracted from transport measurements. Dark blue line denotes the contour for $\epsilon_B = 0$.}
\label{fig5}
\end{figure}

To further investigate the influence of the probing ABS, Fig.~\ref{fig4}(c) illustrates the evolution of the conductance peak height at $[\alpha_+]$ as a function of $\mu_A$. This peak, appearing for positive bias, is proportional to $|v_A|$~\cite{Lange2022} and reaches its maximum value in the linear dispersion region of the ABS. Remarkably, we also observe the emergence of a resonant feature around $\mu_A/\Delta\approx -0.7$ for the chosen parameters, which becomes dominant with increasing coupling strength --  prevalent in the sublinear regime. This feature signals hybridization between the ABS and the subgap state in the KC, appearing when the bias satisfies the resonance condition $eV = \epsilon_A(\mu_A) + \epsilon_B$. This effect is the superconducting analog of spectroscopic resonances found in quantum dot-based probes of Majorana systems~\cite{Deng2016, Prada2017, Clarke2017}. A full analysis of this spectroscopic regime, including its potential as a diagnostic tool, is left for future work.

\section{Limitations of the Method}\label{App_C}

Microscopic details, including the spin degree of freedom or the superconductors mediating the coupling between the dots, are important to understand the behavior of quantum dot-based KCs beyond the idealized Kitaev limit. In particular, the model described in Eq.~\eqref{model_Hamiltonian} allows to describe the effect of finite magnetic fields~\cite{Tanaka2024} and the renormalization effects due to the coupling to the superconductors.

\begin{figure}[t!]
\centering
\includegraphics[width=\columnwidth]{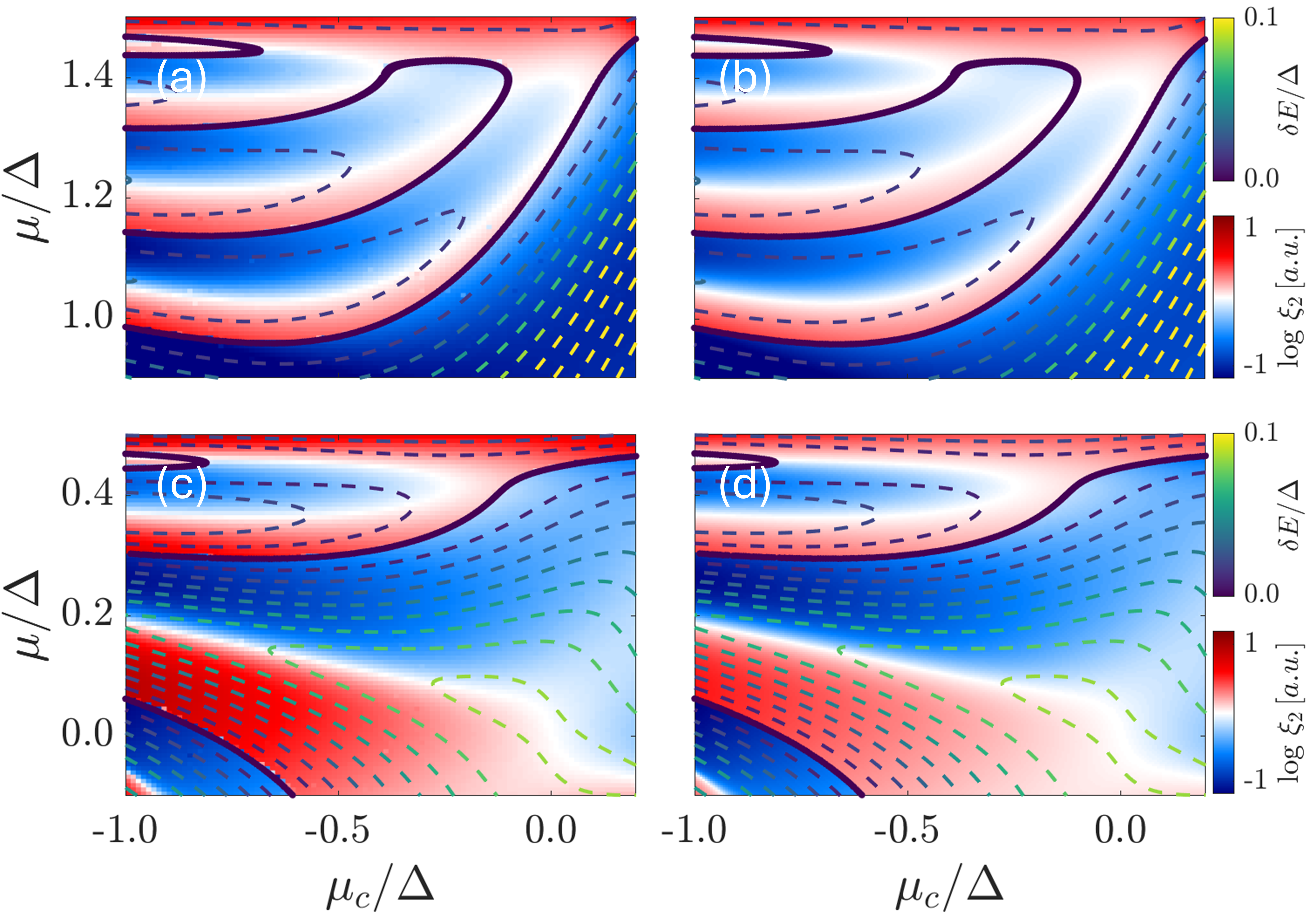}
\caption{Phase diagram of a 5‑site KC for different values of the Zeeman field, $V_z=1.5\,\Delta$ (upper panels), and $V_z=0.5\,\Delta$ (lower panels). Left (Right) column, $\log \, \xi_2$ obtained from transport (diagonalization). Contour dashed lines represent the energy splitting where the dark blue straight line denotes the contour for $\epsilon_B = 0$.}
\label{fig6}
\end{figure}

In this context, it is also important to consider the global Majorana characteristics of the KC. In the main text we mentioned that as the number of QDs increases, a growing disparity emerges between the local Majorana polarization of the low-energy modes in KCs measured at the end, and the non-local MP in Eq.~\eqref{Global_MP}. Figure~\ref{fig5}(a) shows the non-local MP $|{\cal P}|=0$ for a 5-sites KC, see Fig.~3(a) of the main text, that shows the corresponding local MP for comparison. Fig.~\ref{fig5}(b) shows the $\log\,\xi_2$ coefficient, that goes to 0 whenever $u_B=v_B$ for the state in the KC. In other words, while certain sites may exhibit a vanishing local charge, and therefore a potentially high local MP, this does not necessarily imply well-isolated Majorana zero modes. Therefore, there is not a 1 to 1 correspondence between local and non-local MP as these quantities deviate when the system becomes longer and more complex. Nevertheless, the local MP is a good measurement for the local Majorana character of the low-energy states appearing in KCs.

We now focus on the role of the Zeeman field and its impact on the low‑energy subgap states. Figure~\ref{fig6} compares the Bogoliubov-de Gennes (BdG) coherence factors extracted from transport measurements (left panels) with those obtained via direct diagonalization (right panels) of the Hamiltonian for varying strengths of the Zeeman field. In the regime where the system remains within the Kitaev limit $V_z \gtrsim \Delta > t$, shown in Figs.~\ref{fig6}(a,b), there is an excellent agreement between the coherence factors extracted from transport and those obtained by direct diagonalization, validating the accuracy of the transport-based approach. 
As the system approaches the regime $\Delta > V_z \sim t$, Figs.~\ref{fig6}(c,d), where spin polarization is reduced, this agreement remains good, particularly in regions where $|u_B| \sim |v_B|$. This highlights the robustness of our method, to extract local information of the wavefunction.

\begin{figure}[t!]
\centering
\includegraphics[width=\columnwidth]{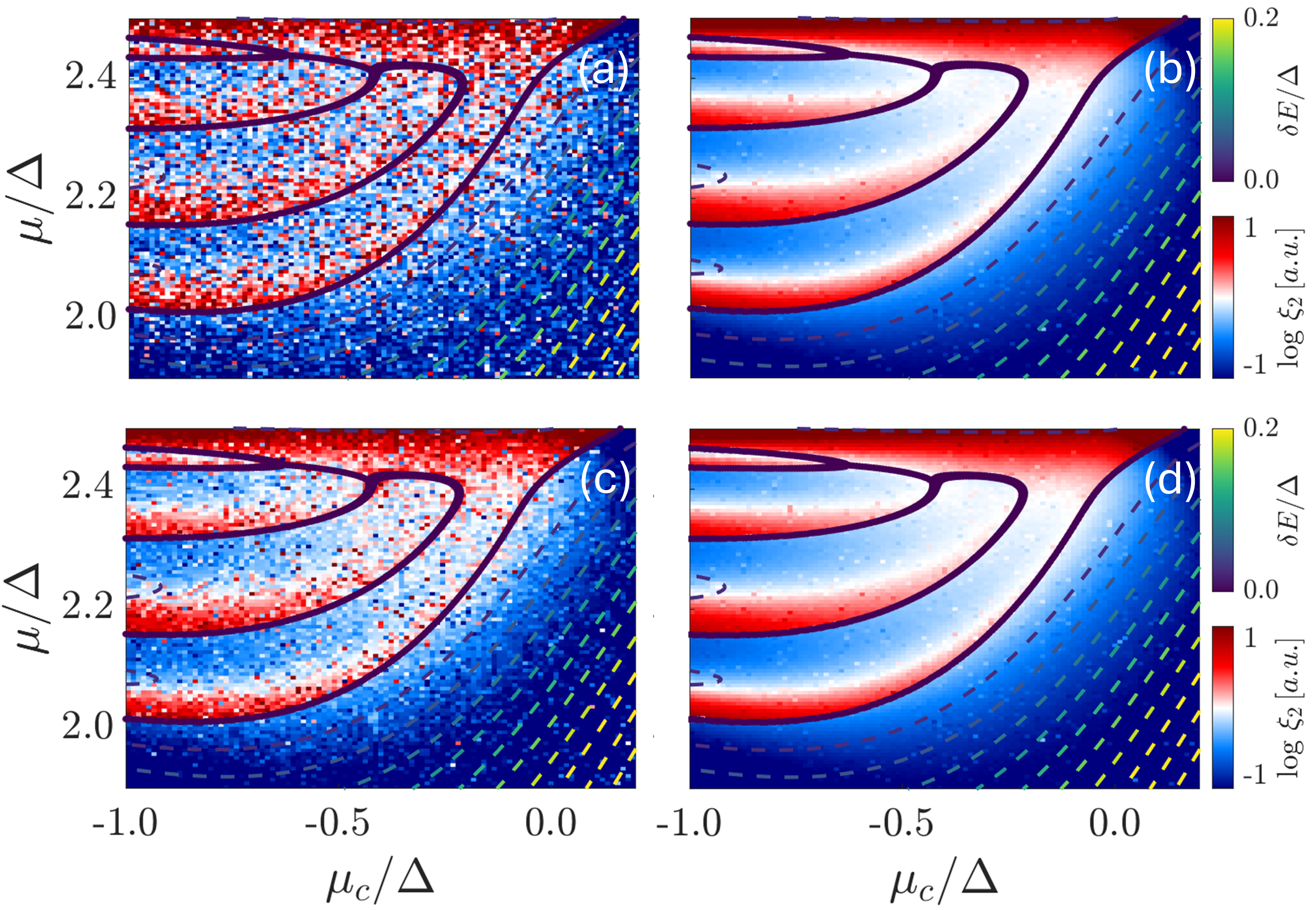}
\caption{Phase diagram of a 5‑site KC under different noise conditions, random (upper panels), and systematic (lower panels). Left (Right) column, noise amplitude $\delta eV/\Gamma_t=1$ ($\delta eV/\Gamma_t=0.1$). Contour dashed lines represent the energy splitting where the dark blue straight line denotes the contour for $\epsilon_B = 0$.}
\label{fig7}
\end{figure}

To further assess the robustness of our approach, we examine the impact of voltage bias fluctuations -- specifically, systematic noise (uniform shifts across all peaks) and random noise (independent shifts with stochastic magnitudes and signs). As shown in Fig.~\ref{fig7}, the method remains accurate as long as the noise amplitude $\delta eV$ does not exceed the broadening of the peaks, which is set by the dominant relaxation rate, {\it i.e.}, $\Gamma_t$ in the linear regime. Within this range, the spectroscopic features of the subgap states remain well resolved, allowing for reliable extraction of the BdG coherence factors.

A closer analysis of each noise source reveals distinct behaviors. Figs.~\ref{fig7}(a,b) show that random noise generally leads to reduced performance and slower convergence as the noise amplitude increases. Conversely, Figs.~\ref{fig7}(c,d) show that the method is significantly more resilient to systematic voltage fluctuations affecting to all peaks in the same way. The phase diagram when $\delta eV/\Gamma_t\gtrsim1$ closely mirrors that obtained under reduced splitting $\epsilon_B$ resolution, with noise amplitude now setting the resolution limit rather than thermal broadening (not shown). This supports the validity of approximating the conductance at the threshold voltages using the corresponding peak values, {\it i.e.}, $G_s([\alpha_\pm]) \to \alpha_\pm$ and $G_s([\beta_\pm]) \to \beta_\pm$. Altogether, these results emphasize the reliability of the method under experimentally relevant conditions where fluctuations and noise are unavoidable.

\section{Green's Function Formalism}\label{App_D}

In the main text, we have described a hybrid setup consisting in a quantum-dot based minimal KC coupled to a normal lead and an ABS. We describe the ABS as appearing  in a quantum dot strongly coupled to a superconducting lead, that can be described by the advanced boundary Green's functions (bGF)~\cite{Cuevas1996, Zazunov2016, Alvarado2020, Alvarado2022} written in the $4\times4$ Nambu basis (hat notation) $\hat{\Psi} = (\psi_\up, \psi_\up^\dagger, \psi_\dw, \psi_\dw^\dagger)^T$,
\be
\hat{{\cal G}}^a_n = \frac{i}{t_n}\sigma_0\tau_0 \ , \quad \hat{{\cal G}}^a_s = -\frac{\omega \, \sigma_0\tau_0 - \Delta_s \, \sigma_y\tau_y}{t_s \sqrt{\Delta_s^2-\omega^2}} \, ,
\ee 
where $\tau_\mu$ ($\sigma_\mu$) are Pauli matrices acting in spin (electron/hole) space. We consider the wide-band approximation, ({\it i.e.}, $t_n,\, t_s \gg \omega, \,\Delta_s$) with the coupling terms with the different leads across the junction take the form $\hat{\Sigma}_\mu = \lambda_\mu \, \sigma_z\tau_0$. 

For the advanced Green's functions we are implicitly assuming $\omega \rightarrow \omega - i\eta$ with $\eta \rightarrow 0^+$~\cite{Dynes1978, Bauer2007, Zazunov2016}, and we omit the super‑index $a$ for convenience. 
Following Refs.~\cite{Bauer2007, Zazunov2016}, we define the brunch cut in the square root along the negative axis,
\be
\sqrt{\Delta_s^2-(\omega-i\eta)^2}= \left\lbrace
\begin{array}{cl}
\sqrt{\Delta_s^2-\omega^2}  & |\omega|\leq \Delta_s \, , \\
i \, \textrm{sgn}(\omega)\sqrt{\omega^2-\Delta_s^2}  & |\omega| > \Delta_s \, .
\end{array}
\right.
\ee

The bGF associated with the probing ABS can be expressed in an analytical form~\cite{Bauer2007}
\begin{widetext}
\be
\hat{\mathcal{G}}_L = \left [\omega \, \hat{\mathbb{I}} - \mu_A \, \sigma_0\tau_z - \hat{\Sigma}_s^\dag \, \big [ \hat{{\cal G}}_s \big ]^{-1} \hat{\Sigma}_s \right]^{-1} =  \frac{\omega \, \big (\Gamma_s +\sqrt{\Delta_s^2-\omega^2} \, \big)  \, \sigma_0\tau_0 + \mu_A \sqrt{\Delta_s^2-\omega^2} \, \sigma_0\tau_z + \Delta_s \Gamma_s \, \sigma_y\tau_y}{2  \Gamma_s \, \omega^2 + \sqrt{\Delta_s^2-\omega^2} \, \big [(\omega^2-\mu_A^2)-\Gamma_s^2 \big]} \, ,
\ee   
\end{widetext}
where the poles of the bGF are contained in the characteristic polynomial~\cite{Bauer2007} from which it is possible to obtain $\epsilon_A$. To keep the presentation concise, we omit the full expression.

To compute the bGF at the edge of a general chain, we adopt the recursive approach of Ref.~\cite{Alvarado2022}. In this scheme, the bGF at the $n$‑th site is expressed as
\begin{subequations}
\begin{gather}
\big [ \hat{\mathcal{G}}_R(n) \big]^{-1} = \omega \, \hat{\mathbb{I}} - \hat{H}_0(n)  - \hat{\Sigma}_R(n) \, ,  \\
\hat{\Sigma}_R(n) = \hat{H}_T \, \big [ \hat{\mathcal{G}}_R(n-1) \big]^{-1}\hat{H}_T^\dag \, ;
\end{gather} 
\end{subequations}
where $\hat{H}_0(n)$ denotes the local Hamiltonian of the $n$‑th site, and $\hat{\Sigma}_R(n)$ is the self‑energy that accounts for the coupling between sites $n$ and $n-1$ via the hopping term $\hat{H}_T$. This formulation enables an efficient, site‑by‑site construction of the bGF for a finite chain. The spinful Hamiltonian in this basis follows

\begin{subequations} 
\bea 
&\hat{H}_0(2k+1) =  \mu \, \sigma_0\tau_z  + V_z \,\sigma_z\tau_z \, ,& \\
&\hat{H}_0(2k) =  \mu_c \, \sigma_0 \tau_z - \Delta \, \sigma_y\tau_y  \, ,& \\
&\hat{H}_T = t \, \sigma_0 \tau_z - i \, t_{so} \, \sigma_y \tau_z \, ;
\eea
\end{subequations}
where $k = {0, \dots, n-1}$ for an $n$‑site chain. It should be noted that the total length of the KC, $L = 2n - 1$, must be an odd positive integer to ensure well-defined boundaries. 

In the simplified case of an isolated minimal KC ($n=2$), the Green's function of the system can be computed analytically. This allows one to derive the characteristic polynomial of the system and determine the conditions under which zero-energy modes emerge at the chain’s boundaries, given by
\be 
\mu^\pm_c=\frac{2\mu(t^2+t_{so}^2)\pm\sqrt{4 V_z^2 (t^2-t_{so}^2)^2 - \Delta^2 (\mu^2-V_z^2)^2}}{(\mu^2-V_z^2)} \, .
\ee

\onecolumngrid
\section{Current Calculation}\label{App_E}

We first study the tunneling between an ABS and a spinless superconducting subgap state. A convenient approach to analyze voltage-biased transport in superconducting hybrid systems is to adopt a gauge where the chemical potential difference appears as a time-dependent tunnel coupling. In this framework, the superconducting phase difference evolves as $\phi(t)=\phi_0 + (2\,eV/\hbar)\, t$, cf. Ref.~\cite{Zazunov2016}. 

To gain analytical insight, we focus on the weak-coupling regime, where Andreev reflections in the probe can be neglected, {\it i.e.}, the anomalous component of the probe can be disregarded. It should be noted that MAR, that represent higher-order tunneling processes, are strongly suppressed for small couplings between the states. In particular, for a MZM the probing ABS couples only to a single spin species, fully suppressing MAR~\cite{Zazunov2012, Peng2015, Zazunov2016, Alvarado2024}. 
Nevertheless, Andreev transport into the target subgap state persists, being resonantly enhanced due to the presence of the latter. As a result, the current through a probing ABS and the subgap state, expressed in the $2\times2$ Nambu basis $\Psi = (\psi_\up, \psi_\up^\dagger)^T$, takes the form, 
\be \label{eq_curr}
I = \frac{e \lambda}{h} \, \Re \int \frac{1}{2} \, \textrm{tr}_N \left \{ G^{-+}_{RR}(\omega)\Sigma^\dag \bar{{\cal G}}_L^{+-}(\omega) - G^{+-}_{RR}(\omega)\Sigma^\dag \bar{{\cal G}}_L^{-+}(\omega) \right \}d\omega \, .
\ee 
To account for an unpolarized spinful subgap state instead ({\it e.g.}, when considering KCs in the main text), an extra factor of 2 must be included in the current to reflect the full spin structure of the state~\cite{Peng2015}. Additionally, we have used the commutation properties of the trace as both, the non-equilibrium bGF for the probing ABS without the anomalous part $\bar{{\cal G}}_L$, and the coupling term between subgap states $\Sigma = \lambda \sigma_z$ are diagonal in Nambu space.

It should be noted that the probing ABS effectively behaves as a gaped normal lead~\cite{Perrin2022}, as the anomalous part are neglected, and the voltage bias in the the non-equilibrium bGF enters as
\begin{subequations}
\begin{gather}
\bar{{\cal G}}^{+-}_L(\omega) = F(\omega) \big [ {\cal G}_L^a - {\cal G}_L^r\big] = 2\pi i \, \textrm{diag} \big [ n_F(\omega_-)\rho_{A,e}(\omega_-), \, n_F(\omega_+)\rho_{A,h}(\omega_+) \big ] \, , \\
\bar{{\cal G}}^{-+}_L(\omega) = -(\mathbb{I} -F(\omega)) \big [ {\cal G}_L^a - {\cal G}_L^r\big] = -2\pi i \, \textrm{diag} \big [ (1-n_F(\omega_-)) \rho_{A,e}(\omega_-), \, (1-n_F(\omega_+)) \rho_{A,h}(\omega_+) \big] \, ;
\end{gather}
\end{subequations}
being $F(\omega) = \diag[n_F(\omega_-),n_F(\omega_+)]$ the quasi-equilibrium distribution functions, where $n_F(\omega)$ the Fermi-Dirac distribution function, and $\omega_\mp = \omega \mp eV$. We define the local spectral density associated to the probing ABS state as $\rho_{A,e/h} (\omega) = \Im \, \bar{{\cal G}}_{L,e/h} (\omega)/\pi$. 

To compute the non-equilibrium bGF of the target subgap state including the effects associated to the coupling with the ABS we use Langreth rules~\cite{Langreth1972, Langreth1976, Cuevas1999} ({\it i.e.}, $[AB]^{+-}=A^rB^{+-} + A^{+-}B^a$ where the  super-index $r/a$ denotes the retarded/advanced GFs) when considering the self-energy $\Sigma_R = \Sigma^\dag \bar{{\cal G}}_L \Sigma$,
\be \label{non_eq_GR}
G^{+-/-+}_{RR} = \big [{\cal G}_R + {\cal G}_R \Sigma_R G_{RR} \big ]^{+-/-+} = \big [ \mathbb{I} - {\cal G}_R^r \Sigma_R^r \big ]^{-1} \big [ {\cal G}_R^{+-/-+} (\mathbb{I} + \Sigma_R^a G_{RR}^a) + {\cal G}_R^r \Sigma_R^{+-/-+} G_{RR}^a \big ] \, ,
\ee
Considering a small splitting $\epsilon_B \rightarrow 0$ for the target subgap mode~\cite{Peng2015} we define the bare bGF for the latter as
\be 
{\cal G}_R^{r,a} = \frac{1}{(\omega-\epsilon_B) \pm i\, \Gamma_t/2} \bmat g_{ee} && g_{eh} \\ g_{he} && g_{hh} \emat \, ,
\ee 
and the non-equilibrium GFs considering that the unperturbed bGFs in this Gauge are in thermodynamic equilibrium, thus
\begin{subequations}
\begin{gather}
{\cal G}_R^{+-} = n_F(\omega) \big [ {\cal G}_R^a - {\cal G}_R^r\big] = \frac{i \, \Gamma_2 }{(\omega-\epsilon_B)^2 + \Gamma_t^2/4} \bmat g_{ee} && g_{eh} \\ g_{he} && g_{hh} \emat \, , \\
{\cal G}_R^{-+} = -(1-n_F(\omega)) \big [ {\cal G}_R^a - {\cal G}_R^r\big] = \frac{- i \, \Gamma_1 }{(\omega-\epsilon_B)^2 + \Gamma_t^2/4} \bmat g_{ee} && g_{eh} \\ g_{he} && g_{hh} \emat \, .
\end{gather}
\end{subequations}
Following Ref.~\cite{Ruby2015}, we have considered the relaxation processes, $\Gamma_1 = \Gamma_t \, [1-n_F(\omega)]$ for emptying, and $\Gamma_2 = \Gamma_t \, n_F(\omega)$ for filling the state. Moreover, the total thermal broadening $\Gamma_t=\Gamma_1+\Gamma_2$ could be related to both the temperature and the coupling to the normal lead~\cite{Ruby2015, Gramich2017, Gorm2022}. 

Using the Dyson equation, we obtain the ``dressed'' bGF of the target subgap mode $G_{RR} = [\mathbb{I}-{\cal G}_R \Sigma_R ]^{-1}{\cal G}_R$ valid for both retarded and advanced components. By taking advantage of the diagonal structure of $\Sigma_R$, and defining $\tilde{\omega}=(\omega-\epsilon_B \pm i\, \Gamma_t/2)$, we get
\be 
G_{RR}^{r,a} = \frac{\bmat g_{ee} \, \tilde{\omega} \pm i \pi \lambda^2 \rho_{A,h}(\omega_+) \det g  && g_{eh} \, \tilde{\omega} \\ g_{he} \, \tilde{\omega} && g_{hh} \, \tilde{\omega} \pm i \pi \lambda^2 \rho_{A,e}(\omega_-) \det g \emat}{\tilde{\omega}^2 - \lambda(\omega) \pm i \pi \tilde{\omega} \lambda^2 \big [g_{ee} \, \rho_{A,e}(\omega_-) + g_{hh} \, \rho_{A,h}(\omega_+)\big ]} \, ,
\ee
where $\lambda(\omega) = \pi^2 \lambda^4 \rho_{A,e}(\omega_-) \rho_{A,h}(\omega_+) \det g$, being $\det g = g_{ee}\,g_{hh}-g_{eh}\,g_{he}$, and we have considered just the imaginary part of the self energy $\Sigma_R^{r/a} = \mp i \pi \lambda^2 \textrm{diag} [\rho_{A,e}(\omega_-), \rho_{A,h}(\omega_+)]$ which leads to the frequency dependent tunneling rates,
\be 
\Gamma_e(\omega_-) = 2\pi\lambda^2 g_{ee} \,\rho_{A,e}(\omega_-) \, , \qquad \Gamma_h(\omega_+) = 2\pi\lambda^2 g_{hh} \, \rho_{A,h}(\omega_+) \, .
\ee 
We note that the preceding formulas provide some corrections of those presented in Ref.~\cite{Peng2015} for the case of a BCS probing a subgap state.

\section{Analytical Relations for the Current}\label{App_F}

Since we focus on the linear regime~\cite{Peng2015, Ruby2015, Alvarado2024} ({\it i.e.}, weak coupling where relaxation is faster than tunneling $\Gamma_t \gg \Gamma_{e/h}$), potential contributions to the current arising from $\det g \propto \lambda^4$ can be neglected. These terms should be regarded as higher-order corrections accounted in the full computational treatment when using Eq.\eqref{eq_curr} and the bGF in the previous section. Furthermore, for small detunings in the Kitaev limit, we recover an ideal partially splitted MZM $\gamma_B = u_B \, c + v_B \, c^\dag$, characterized by $\det g=0$, satisfying
\be \label{PMM_GF_An}
G_{RR}^{r,a} = \frac{1}{\omega-\epsilon_B \pm i\, \Gamma(\omega)/2} \bmat |u_B|^2 && u_B \, v_B^* \\ u_B^* \, v_B && |v_B|^2 \emat,
\ee
where $\Gamma(\omega) = \Gamma_e(\omega_-) + \Gamma_h(\omega_+) + \Gamma_t$ is the total broadening. Substituting in Eq.\eqref{non_eq_GR} for the non-equilibrium bGF we obtain
\begin{subequations}
    \begin{gather}
        G^{+-}_{RR} = i \, \frac{\Gamma_2 + \Gamma_e(\omega_-) \, n_F(\omega_-)+\Gamma_h(\omega_+) \, n_F(\omega_+)}{(\omega-\epsilon_B)^2 + (\Gamma_t+\Gamma_e(\omega_-)+\Gamma_h(\omega_+))^2/4} \bmat |u_B|^2 && u_B \, v_B^* \\ u_B^* \, v_B && |v_B|^2 \emat, \\
        G^{-+}_{RR} = -i \, \frac{\Gamma_1 + \Gamma_e(\omega_-)(1-n_F(\omega_-))+\Gamma_h(\omega_+)(1-n_F(\omega_+))}{(\omega-\epsilon_B)^2 + (\Gamma_t+\Gamma_e(\omega_-)+\Gamma_h(\omega_+))^2/4} \bmat |u_B|^2 && u_B \, v_B^* \\ u_B^* \, v_B && |v_B|^2 \emat.
    \end{gather}
\end{subequations}

Substituting in Eq.\eqref{eq_curr} we obtain the total spin polarized current in Ref.~\cite{Ruby2015, Peng2015} modified to account for the ABS probe, where the Andreev and single-particle contributions $I=I_A+I_{sp}$ take the form
\begin{subequations}
    \begin{gather}
        I_A(V) = \frac{e}{h} \int \frac{\Gamma_e(\omega_-)\,\Gamma_h(\omega_+) \, [n_F(\omega_-)-n_F(\omega_+)]}{(\omega-\epsilon_B)^2 + (\Gamma_t+\Gamma_e(\omega_-)+\Gamma_h(\omega_+))^2/4} \, d\omega \, , \\ \nonumber \\
        I_{sp}(V) = \frac{e}{2h} \int \frac{\Gamma_1 \,[\Gamma_e(\omega_-) \, n_F(\omega_-)-\Gamma_h(\omega_+) \, n_F(\omega_+)] -\Gamma_2 \, [\Gamma_e(\omega_-)(1-n_F(\omega_-))-\Gamma_h(\omega_+)(1-n_F(\omega_+))]}{(\omega-\epsilon_B)^2 + (\Gamma_t+\Gamma_e(\omega_-)+\Gamma_h(\omega_+))^2/4} \, d\omega \, .
    \end{gather}
\end{subequations}

Our aim is to develop a low energy analytical description of the main subgap contributions to the current $e|V|\leq \Delta$ ({\it i.e.}, voltage bias around the Fermi energy) at weak couplings. Therefore, we consider the ABS probing the system in the atomic limit linearized around $|\omega|=\epsilon_A+\delta \omega$~\cite{Bauer2007}
\begin{subequations}
\begin{gather}
|u_A|^2 = \frac{\epsilon_A+\mu_A}{2 \epsilon_A} \, , \quad  |v_A|^2 = \frac{\epsilon_A-\mu_A}{2 \epsilon_A} \, , \\ 
\bar{{\cal G}}_L(\omega) \approx \frac{1}{\omega-\epsilon_A} \bmat |u_A|^2 && 0 \\ 0 && |v_A|^2 \emat + \frac{1}{\omega+\epsilon_A} \bmat |v_A|^2 && 0 \\ 0 && |u_A|^2 \emat ;
\end{gather}
\end{subequations}
where the local spectral density associated to the ABS satisfies $\rho_{A,e}(\omega)=\rho_{A,h}(-\omega)$, with
\be \label{DoS_ABS}
\rho_{A,e/h}(\omega) = \frac{\Lambda}{\pi} \left[\frac{|u_A|^2}{(\omega\mp\epsilon_A)^2+\Lambda^2} + \frac{|v_A|^2}{(\omega\pm\epsilon_A)^2+\Lambda^2} \right] \, .
\ee
Here, $\Lambda$ is a small broadening term that can be associated to the quasiparticle lifetime of the probing ABS, giving the characteristic width of the spectral weight features~\cite{Dynes1978}. Typical values for this parameter in conventional superconductors are $\Lambda < 10^{-2} \cdot \Delta$~\cite{Cuevas1999}. We assume $\Lambda \leq \Gamma_t$ to remain within experimentally realistic broadening scales.

We consider the different contributions to the single particle current selecting the linear part of the ABS such $|u_A|^2 \approx 1$, thus $\Gamma_e \gg \Gamma_h$ where for negative voltages, $1-n_F(\omega-eV) = 1-n_F(\omega + e|V|) \approx n_F(\omega-eV) \approx 1$. Analogously, for $|v_A|^2 \approx 1$, thus $\Gamma_h \gg \Gamma_e$ where for negative voltages, $n_F(\omega+eV) = n_F(\omega-e|V|) \approx 1-n_F(\omega+eV) \approx 1$. After some algebra, the main contributions to the current following the linear dispersion of the ABS induced at the threshold $e|V|=\epsilon_A+\epsilon_B$ satisfies
\begin{subequations} 
\bea \label{sp_current}
I_{sp}(V) &=& \frac{e}{2h} \int \frac{\Gamma_1 \, \Gamma_e(\omega_-) \, n_F(\omega_-) + \Gamma_2 \, \Gamma_h(\omega_+)(1-n_F(\omega_+))}{(\omega-\epsilon_B)^2 + (\Gamma_t+\Gamma_e(\omega_-)+\Gamma_h(\omega_+))^2/4} \, d\omega \, , \\
I_{sp}(-V) &=& -\frac{e}{2h} \int \frac{\Gamma_2 \, \Gamma_e(\omega_-)(1-n_F(\omega_-))+\Gamma_1 \, \Gamma_h(\omega_+) \, n_F(\omega_+)}{(\omega-\epsilon_B)^2 + (\Gamma_t+\Gamma_e(\omega_-)+\Gamma_h(\omega_+))^2/4} \, d\omega \, .
\eea
\end{subequations}
Retaining the relevant poles in $\Gamma_e$ and $\Gamma_h$ from Eq.\eqref{DoS_ABS} in the integration of the current at the threshold 
\begin{subequations} 
\bea \label{main_sp_curr}
I_{sp}(V) &\approx& \frac{e \lambda^2}{h} \int \frac{\Lambda \, |v_A(\mu_A)|^2}{(\omega-\epsilon_B)^2 + \Gamma_t^2/4} \left[ \Gamma_1 \frac{|u_B|^2}{(\omega_-+\epsilon_A)^2+\Lambda^2}+\Gamma_2 \frac{|v_B|^2}{(\omega_+-\epsilon_A)^2+\Lambda^2} \right] \, d\omega \, , \\
I_{sp}(-V) &\approx& - \frac{e\lambda^2}{h} \int \frac{\Lambda \, |u_A(\mu_A)|^2}{(\omega-\epsilon_B)^2 + \Gamma_t^2/4} \left [ \Gamma_2 \frac{|u_B|^2}{(\omega_--\epsilon_A)^2+\Lambda^2}+\Gamma_1 \frac{|v_B|^2}{(\omega_++\epsilon_A)^2+\Lambda^2} \right] \, d\omega \, .
\eea
\end{subequations}
Notice that, there are generally mixed contributions from the BdG coherence factors of the MZMs proportional to $\Gamma_1$ and $\Gamma_2$. 

\section{Analytical Relations for the Conductance}\label{App_G}

The differential conductance provides a more robust way to extract the BdG coherent factors of the target state, circumventing some of the challenges, like the the aforementioned mixed contributions. Therefore, it allows to separate the contributions arising from $\Gamma_1$ and $\Gamma_2$.

To gain analytical insight into the conductance, we consider Eq.\eqref{sp_current} in the linear regime at zero temperature. In this limit, the Fermi distribution simplifies to $n_F(\omega_\pm) = [1-\Theta(\omega_\pm)]$, with $\Theta(\omega_\pm)$ denoting the Heaviside step function, and its voltage derivative becomes $\partial \, n_F(\omega_\pm)/\partial V = \mp e \delta(\omega_\pm)$. 
The main contributions to the conductance at positive voltage bias from Eq.\eqref{main_sp_curr} are obtained as

\begin{subequations}
\bea
i) \quad 2\int_0^\infty  && \frac{ n_F(\omega_-)}{(\omega-\epsilon_B)^2 + \Gamma_t^2/4}  \frac{\Lambda \, |u_B|^2 \, |v_A|^2}{(\omega_-+\epsilon_A)^2+\Lambda^2} \, d\omega  =
\nonumber \\
&& \frac{2 \, \Lambda \, |u_B|^2 \, |v_A|^2}{\Gamma_t/2 [(eV-\epsilon_B-\epsilon_A)^2+(\Gamma_t/2-\Lambda)^2][(eV-\epsilon_B-\epsilon_A)^2+(\Gamma_t/2+\Lambda)^2]} \times 
\nonumber \\
&& \Bigg \{ \Gamma_t/2 \Lambda (eV-\epsilon_B-\epsilon_A) \log \left( \frac{[(eV-\epsilon_B)^2+\Gamma_t^2/4][(eV-\epsilon_A)^2+\Lambda^2]}{(\Lambda^2+\epsilon_A^2)(\Gamma_t^2/4+\epsilon_B^2)} \right) + 
\nonumber \\
&& \Gamma_t/2 \left [(eV-\epsilon_B-\epsilon_A)^2+(\Gamma_t^2/4-\Lambda^2) \right] \left [\mathrm{atan} \left(\frac{eV-\epsilon_A}{\Lambda}\right ) + \mathrm{atan} \left(\frac{\epsilon_A}{\Lambda}\right ) \right] + 
\nonumber \\
&& \Lambda \left [(eV-\epsilon_B-\epsilon_A)^2-(\Gamma_t^2/4-\Lambda^2) \right] \left [ \mathrm{atan} \left(\frac{eV-\epsilon_B}{\Gamma_t/2} \right) + \mathrm{atan} \left(\frac{\epsilon_B}{\Gamma_t/2}\right )\right]
\Bigg \} \, , \label{Ge_1} 
\eea

\bea
ii) \quad 2\int_{-\infty}^0  && \frac{ n_F(\omega_+)}{(\omega-\epsilon_B)^2 + \Gamma_t^2/4} \frac{\Lambda \, |v_B|^2 \, |v_A|^2}{(\omega_+-\epsilon_A)^2+\Lambda^2} \, d\omega =
\nonumber \\
&& \frac{2 \, \Lambda \, |u_B|^2 \,|v_A|^2}{\Gamma_t/2 [(eV+\epsilon_B-\epsilon_A)^2+(\Gamma_t/2-\Lambda)^2][(eV+\epsilon_B-\epsilon_A)^2+(\Gamma_t/2+\Lambda)^2]} \times 
\nonumber \\
&& \Bigg \{ \Gamma_t/2 \Lambda (eV+\epsilon_B-\epsilon_A) \log \left( \frac{[(eV+\epsilon_B)^2+\Gamma_t^2/4][(eV-\epsilon_A)^2+\Lambda^2]}{(\Lambda^2+\epsilon_A^2)(\Gamma_t^2/4+\epsilon_B^2)} \right) + 
\nonumber \\
&& \Gamma_t/2 \left [(eV+\epsilon_B-\epsilon_A)^2+(\Gamma_t^2/4-\Lambda^2) \right] \left [\mathrm{atan} \left(\frac{eV-\epsilon_A}{\Lambda} \right ) + \mathrm{atan} \left(\frac{\epsilon_A}{\Lambda} \right ) \right] + 
\nonumber \\
&& \Lambda \left [(eV+\epsilon_B-\epsilon_A)^2-(\Gamma_t^2/4-\Lambda^2) \right] \left [ \mathrm{atan} \left(\frac{eV+\epsilon_B}{\Gamma_t/2} \right ) - \mathrm{atan} \left(\frac{\epsilon_B}{\Gamma_t/2} \right ) \right]
\Bigg \} \, . \label{Ge_2}
\eea
\end{subequations}

It can be readily shown that, in this limit, no additional contributions to the conductance arise. As a result, the calculation can be specifically evaluated at the relevant positive voltage thresholds, $eV=\epsilon_A\pm\epsilon_B$,
\begin{subequations}
\bea 
G_{sp}([\alpha_+]) = \frac{e^2\lambda^2}{h} \frac{\Lambda \, \Gamma_{1} \, |u_B|^2 \, |v_A|^2}{(\Gamma_t^2/4-\Lambda^2)^2}\Bigg [\log \left( \frac{(\Gamma_t^2/4+\epsilon_A^2)(\Lambda^2+\epsilon_B^2)}{(\Lambda^2+\epsilon_A^2)(\Gamma_t^2/4+\epsilon_B^2)}\right) + (\Gamma_t^2/4-\Lambda^2) \left(\frac{1}{\Lambda^2 + \epsilon_B^2} -\frac{1}{\Gamma_t^2/4 + \epsilon_A^2} \right) \Bigg] \, , 
\nonumber \\  \\
G_{sp}([\beta_+]) = \frac{e^2\lambda^2}{h} \frac{\Lambda \, \Gamma_{2} \, |v_B|^2 \, |v_A|^2}{(\Gamma_t^2/4-\Lambda^2)^2}\Bigg [\log \left( \frac{(\Gamma_t^2/4+\epsilon_A^2)(\Lambda^2+\epsilon_B^2)}{(\Lambda^2+\epsilon_A^2)(\Gamma_t^2/4+\epsilon_B^2)}\right) + (\Gamma_t^2/4-\Lambda^2) \left(\frac{1}{\Lambda^2 + \epsilon_B^2} -\frac{1}{\Gamma_t^2/4 + \epsilon_A^2} \right) \Bigg] \, . \nonumber \\
\eea
\end{subequations}
The sign of the bias determines the carriers injected, such that for positive bias transport is proportional to the hole components of the probing ABS~\cite{Lange2022}. Determining the conductance at the negative voltage thresholds, $eV=-(\epsilon_A\pm\epsilon_B)$, is a straightforward calculation requiring the transformation $(v_A \rightarrow u_A)$, and $(u_B \leftrightarrow v_B)$. 

We note that finite temperature is required to sustain single-particle current and resolve thermal conductance peaks at $e|V|=\epsilon_A-\epsilon_B$. However, these results can be extended to finite temperature when satisfying $\epsilon_B > k_BT \gg \Gamma_t$~\cite{Perrin2022}, as conductance peak line shapes are renormalized uniformly when peaks remain well separated~\cite{Flensberg2020}. Consequently, the smallest resolvable energy splitting is on the order of the thermal broadening  ($\epsilon_B \gtrsim \Gamma_t$).

Remarkably, by certain arranges of the conductance at the different thresholds, we are able to characterize the junction as

\begin{gather}
\xi_1 = \sqrt{\rule{0pt}{3.2ex} \frac{G_{sp}([\alpha_-]) \, G_{sp}([\beta_-])}{G_{sp}([\alpha_+]) \, G_{sp}([\beta_+])}}= \frac{\big |u_A\big |^2}{\big |v_A\big |^2} \, , \hspace{5em}
\xi_2 = \sqrt{\rule{0pt}{3.2ex} \frac{G_{sp}([\alpha_+]) \, G_{sp}([\beta_-])}{G_{sp}([\alpha_-]) \, G_{sp}([\beta_+])}} = \frac{\big |u_B\big |^2}{\big |v_B\big |^2} \, , \nonumber \\ \nonumber \\
\xi_3 = \sqrt{\rule{0pt}{3.2ex} \frac{G_{sp}([\alpha_+]) \, G_{sp}([\alpha_-])}{G_{sp}([\beta_+]) \, G_{sp}([\beta_-])}} = \frac{\Gamma_1}{\Gamma_2} \, ;
\end{gather}

where we have assumed the notation for the voltage thresholds $[\alpha_\pm]=\pm(\epsilon_A+\epsilon_B)$ and $[\beta_\pm]=\pm(\epsilon_A-\epsilon_B)$. In the spinful case, the conductance depends on spin-resolved BdG coherence factors through the total amplitudes $|u_\nu|^2 = |u_{\nu,\up}|^2+|u_{\nu,\dw}|^2$ and $|v_\nu|^2 = |v_{\nu,\up}|^2+|v_{\nu,\dw}|^2$, for both subgap states $\nu=\{A,B\}$. It should be noted that, in the specified limits, these expressions are exact.

However, the conductance peak does not perfectly coincide with the threshold voltages~\cite{Ruby2015}. Nevertheless, a reasonable approximation can be made by approximating the peak heights to the conductance values at the voltage thresholds, $G_{sp}([\alpha_\pm]) \rightarrow \alpha_\pm$ and $G_{sp}([\beta_\pm]) \rightarrow \beta_\pm$. Thus, from the latter, it is possible to extract not only the coherence factors of both the probing ($\xi_1$) and the target ($\xi_2$) subgap states, but also the thermal broadening ratios ($\xi_3$), 
establishing a direct relation to the system’s thermal properties~\cite{ALY2014, Ruby2015, Cuevas2020b, Gorm2022}.

Finally, we bring attention to the fact that the observables used to extract the BdG coherence–factor ($\xi_1$, $\xi_2$) are insensitive to any smooth energy dependence of the tunnel amplitude $\lambda(\omega)$ in the linear regime, assuming $\alpha_\pm = \lambda^2(\omega_1) F_{1\pm}$, and $\beta\pm = \lambda^2(\omega_2) F_{2\pm}$, where $\omega_{1/2}=\epsilon_A\pm\epsilon_B$. 




\end{document}